# Theoretical investigation on the Schelling's critical neighborhood demand

Jae Kyun Shin[a,*], Hiroki Sayama[b]

[a] School of Mechanical Engineering, Yeungnam University, Kyongsan 712-749, South Korea

[b] Collective Dynamics of Complex Systems Research Group, Binghamton University, State University of New York Binghamton, NY 13902-6000, USA

[*] Correspondence author

ABSTRACT

We derived the critical neighborhood demand in the Schelling's segregation model by studying the conditions for which a chain reaction of migrations of unsatisfied agents occurs. The essence of Schelling dynamics was approximated in two simplified models: (1) a random walk model for the initial stage of the migrations to illustrate the power-law behavior of chain reaction lengths under critical conditions, and (2) a two-room model for the whole process to represent a non-spatial version of segregation dynamics in the Schelling model. Our theoretical results showed good agreements with numerical results obtained from agent-based simulations.

# 1. Introduction

Schelling's segregation model [1-3] is well known in many disciplines, such as sociology, computer science and physics [4]. Despite the debate regarding the validity of the model for explaining the actual segregation phenomena in metropolitan areas, the model is interesting in itself as a typical example of systems that self-organize following very simple rules, and also as one of the first studies that applied what is now called agent-based modeling to social sciences.

Two types of Schelling models are known, the spatial proximity model and the bounded neighborhood model [1,10]. In the spatial proximity model, everyone defines his/her neighborhood by reference to his/her own location. In the bounded neighborhood model, there is a common definition of neighborhood and its boundaries [1]. The present study is mainly concerned with the proximity model defined on a 2D lattice system. Two types of agents were distributed on an 8x8 checkerboard, as an example. Every agent has up to eight neighbors and has a neighborhood demand, such that, for example, at least 1/2 of his/hers neighbors should be of the same type. If the neighborhood demand is below a certain threshold, the system can remain in a well mixed state. If it exceeds critical value, a domino effect of migrations occurs until a striking segregation is obtained as a result. In a hand operated experiment on a small sized checkerboard, Schelling concluded that the critical demand ratio lies between 1/3 and 1/2 [1]. As Schelling originally indicated, the dynamics depend on parameters, such as the neighborhood size (B), neighborhood demand, population ratios and vacancy ratios. Many agent-based models have been reported to determine the effect of these parameters [5-14], to apply the Schelling model for real systems [15-17], or to study the Schelling dynamics on networks [18,19]. Even with those varied models, the key results from the original Schelling model did not change significantly, particularly in terms of the critical demand.

As the original Schelling model is known only in an agent-based format, subsequent analytical studies were performed to explain the Schelling model theoretically. Some of studies suggested that the segregation phenomena of the Schelling model is related to the physics of metal surfaces and crystal growth, and a range of physical models simulating atom or cluster aggregation can be applied to explain the segregation phenomena in the Schelling's model [20]. For example, the Ising model was applied to show the dependence of the critical behavior on a temperature-like parameter [21]. The mean field approximation was used to prove analytically that segregation is the only long-term outcome [22-24]. Evolutionary game theory was applied to prove the existence of the potential function and whether the segregated states are stable [25,26]. Other analytic techniques included methods based on the Lyapunov functions [27] and diffusion equations [28]. Up to the present, analytical methods focused mainly on showing that segregation is a stable outcome when the demand is above the critical value. Agent-based models appear to be the only way of identifying the critical demand, and the key question related to the Schelling dynamics is still waiting to be solved. The present study aims to determine the critical demand theoretically.

## 2. Multiplication factor

Consider two colors of agents on a grid system of size 100x100. Periodic boundaries are assumed. The total number of agents is denoted by N. Assume that the number of agents for each color is the same. The agents have neighborhood demand such that at least T among B of his/hers neighbors should be of the same type. Unlike the original Schelling model, no vacancies were assumed in the present study. Neglecting vacancies simplifies the neighborhood condition and makes it possible to represent the neighborhood demand T with just a single value. Unsatisfied agents migrate in search of a satisfactory neighborhood. Owing to the lack of vacancies, swapping is the only way that the agents can change their locations [27,29]. If there is no vacancy, the size of the neighborhood will be always constant. The present study works with three types of neighborhood topology (B=8, 12 or 24), as shown in Fig. 1.

Initially, the system begins from a well mixed equilibrium state (our method to create such a state will be discussed later). Throughout the present study, an *Equilibrium State* (ES) means any state, in which all agents are satisfied in terms of their neighborhood demand. An ES can be well mixed or highly segregated. Well-mixed ES will be called a *Random Equilibrium State* (RES), which is conceptually similar to the random allocation in reference [24]. In RES, the probability that any given location would be occupied by a particular type equals the ratio of the number of agents of this type to the number of possible locations.

From a RES, an initial disturbance is introduced to the system to start the dynamics. The disturbance is given by swapping a pair of agents at random. Unsatisfied agents can be generated from such initializing compulsory swapping. Whenever unsatisfied agents exist in the system, one is chosen randomly and called $A_1$, and swapped with another agent $A_2$, also chosen randomly. The color of the agents $A_1$ and $A_2$ is denoted as $C_1$ and $C_2$, respectively, but $C_2 \neq C_1$. The group of neighbors of agent A before swapping, will be denoted by H(A). In swapping, unsatisfied agents can be newly generated only among the agents in $H(A_1)$ and $H(A_2)$. The susceptible groups, $G_1$ and $G_2$, among $H(A_1)$ and $H(A_2)$, respectively, denote groups of agents that can become unsatisfied after the migration occurs. For example, group $G_1$ is composed of type $C_1$ agents in $H(A_1)$. Because a type $C_1$ agent moves away, only the $C_1$ type agents in $H(A_1)$ may become unsatisfied.

After the swap, the system may return to the ES or unsatisfied agents may still exist in the system. A series of migrations from an initial compulsory swapping to a new ES will be called a session hereafter. Each session will begin with a compulsory swap at time step $t_0=0$. In every positive time step, the swapping involves an unsatisfied agent. If there are no unsatisfied agents in the system at $t=t_f$, the session ends with a size of $S=t_f-1$. The Schelling dynamics in the present study is interpreted as a series of sessions that repeat until a complete segregation is reached, if ever. The following discussion is valid for time steps $t>0$, for which $A_1$ is always an unsatisfied agent. Assuming that the system is not far from the RES, it can be reasonably assumed that among the members of $H(A_1)$, $G_1$ is composed of $T-1$ agents of color $C_1$. Denoting the size of $G_1$ as $|G_1|$, results in the following:

$$|G_1|=T-1 \qquad (1)$$

Similarly, the susceptible group $G_2$ can be defined as a subgroup of members of $H(A_2)$ with color $C_2$. Assuming that the two sets $H(A_1)$ and $H(A_2)$ do not overlap with each other, the size of the group $G_2$ will be $B/2$ on average:

$$|G_2|=B/2 \qquad (2)$$

The total number of susceptible groups in a swap will be denoted as $|G|=|G_1|+|G_2|$. From the above discussion it follows that

$$|G|=B/2+T-1 \qquad (3)$$

The actual number of unsatisfied agents among groups $G_1$ and $G_2$ will be denoted by $\mu_1$ and $\mu_2$, respectively. Multiplication factor, $\mu\equiv\mu_1+\mu_2$, is defined as the mean number of newly generated unsatisfied agents per swapping. The following equation defines the mean probability, $P$, of a susceptible agent becoming unsatisfied:

$$\mu=|G|P=(B/2+T-1)P \qquad (4)$$

The multiplication factor plays an important role in the behavior of the system. A session is expected to be the longer if $\mu$ is larger. The critical value of multiplication factor can be predicted to be $\mu=1$, as the name itself suggests.

The value of $\mu_1$ and $\mu_2$ can be directly obtained from agent-based simulations, by counting the number of unsatisfied agents among group H($A_1$) and H($A_2$) after swapping. For the numerical simulations, the initial conditions are designed so that the simulation can begin at RES. Typical initial conditions used in the literatures do not satisfy all the conditions of RES. For example, the original checkerboard distribution [1,6] does not satisfy the randomness and completely random initial conditions do not satisfy the equilibrium condition. To obtain an initial RES, we start with a complete checkerboard distribution, which is well mixed and in equilibrium for meaningful values of T, i.e. for T≤B/2. However, it is not random in the sense that the composition of the neighborhood is the same throughout all the agents in the system. To make the distribution random, a pair of agents was swapped one by one, while maintaining the system at equilibrium (the swap is conducted only if it does not destroy the equilibrium condition). For N=10,000, swapping was introduced 5,000 times before the satisfactory initial condition was obtained. The process worked well, resulting in well mixed, random initial conditions in the equilibrium state. In Fig. 2, the multiplication factors obtained from a numerical study are shown for three neighborhood topologies and meaningful ranges of T, respectively. The values from the ABM shown in Fig. 2 were averaged over the first 200 swapping, excluding the initial compulsory ones. As the sessions or time steps increase, the system becomes increasingly segregated and the RES assumption may no longer be valid. For this reason, the multiplication factors were measured using earlier time frames.

## 3. Theoretical evaluation of multiplication factor

One of the main targets of this study was to calculate the multiplication factor theoretically, without resorting to the full-scale ABM. To determine the multiplication factor, several simplifying assumptions were also used, as already explained. Under the given assumptions, the multiplication factors $\mu_1$ and $\mu_2$ could be calculated based on the combinatorial probabilities. Equivalently, a small scale Monte Carlo simulation can be used for the same purpose. A brief description of the small scale Monte Carlo simulations is as follows. Consider the case of B=8 as an example. A 5x5 cell system was used. With the focal agent $A_1$ located at the center, the 5x5 system includes all the members of H($A_1$) and their neighbors. All the cells were filled randomly with a probability of 0.5 for each of the two colors. Only the cases satisfying condition of Eq. (1) were counted as an effective trial. For the effective trial, the color of the center cell was changed and the number of unsatisfied agents was counted as contributing to the multiplication factor $\mu_1$. The multiplication factor $\mu_1$ was obtained by averaging the number of unsatisfied agents with respect to the number of effective trials. Using condition (2), $\mu_2$ can be obtained in a similar manner. Figure 2 also shows the multiplication factors calculated in this way. From Fig. 2, the theoretical value matched the numerical results well.

Depending on the multiplication factors given in Fig. 2, the critical multiplication factor occurs between T=2 and 3, for the case of B=8. This agrees in general with the Schelling's conclusion and other previous numerically determined results in the literature [19, 22]. The multiplication factor can change even for a given simulation, depending on the time step at which the mean value is evaluated. Fig. 3 presents several example cases showing the change in the multiplication factor as a cumulative average. In most cases, initial intervals can be found (t=0~300) for which the cumulative average is almost constant. The probabilistic approach can be valid at least for the initial stage of the system. In addition, for many cases, the change in the multiplication factors is such that they can be assumed to be constant for a longer time step. The results from Figs.2 and 3 show that the theoretical method described in the present study works well.

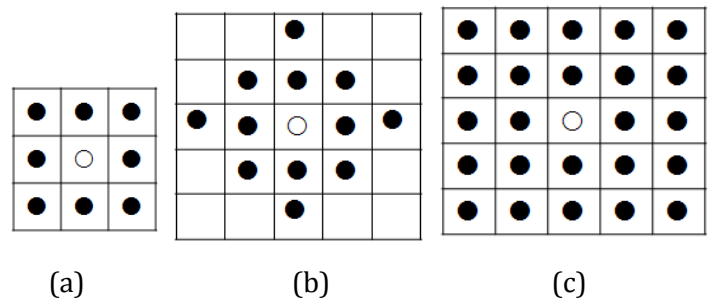

**Fig. 1.** Neighborhood topologies, (a)Moor Neighborhood(B=8),(b)B=12, (c)B=24.

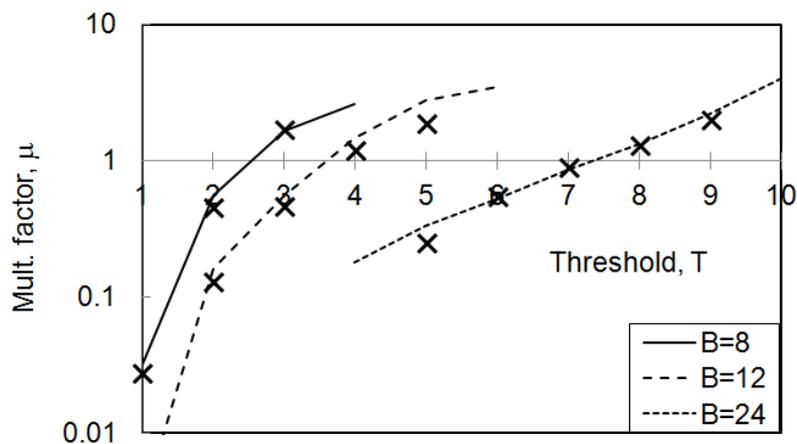

**Fig. 2.** Multiplication factors. Symbols are for the numerical results as obtained from averaging 20 MC simulations per each data point. Theoretical results are shown in lines.

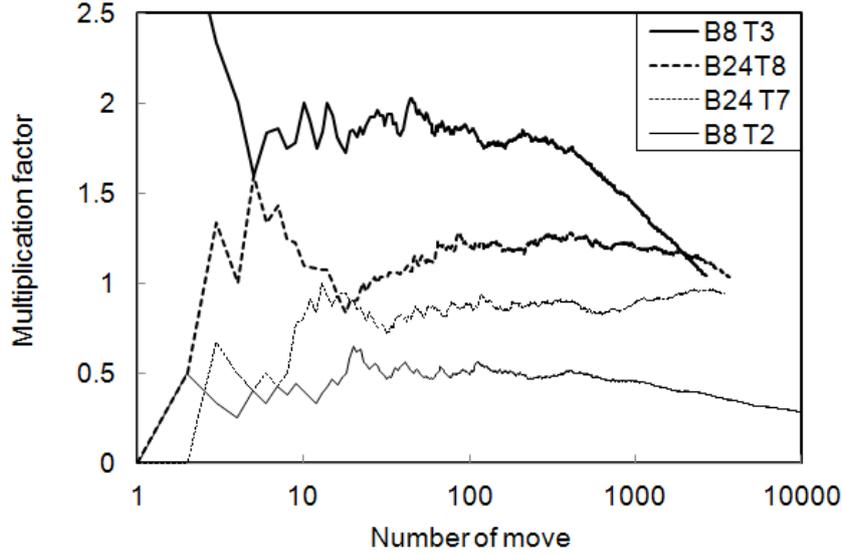

**Fig. 3.** Sample cumulative averages of the multiplication factor μ from the ABM. Cumulative average at time t is the average over the time period between time 0 and t.

## 4. Random walk in Schelling dynamics

From the multiplication factor, a phase transition between T=2 and T=3 can be predicted for the case of B=8. Because the neighborhood demand T is not a continuous value, it does not make sense when used to discuss the precise phase transition point, which should lie between 2<T<3. On the other hand, it is of theoretical interest to predict the behavior of the system for intermediate values of T. For this purpose, the Schelling model was converted to a random walk model [30] in the present study. The mean probability P appearing in Eq.(4) was used. From the result shown in Fig.3, the mean probability $P=\mu/|G|$ can be assumed to be constant for a given simulation, at least for earlier time frames. The total number of unsatisfied agents in the entire system at time step t, denoted by $U_t$, depends on the following random walk. (For details, see Appendix A)

$$U_{t+1} = \begin{cases} U_t + \mu - P, & \text{with probability } P \\ U_t - P, & \text{with probability } 1 - P \end{cases} \quad (5)$$

The step size of the random walk depends on the direction of the move, and the choice of direction depends on the probabilities. The mean step size can be calculated easily as P(μ-1). If μ>1 (or μ<1), the mean step size will be positive (negative) and the number of unsatisfied agents is expected to increase (decrease) with increasing time step. The random walk represented by Eq.(5) has an absorbing barrier at the origin. Starting from $U_0=0$, a survival session for the random walk corresponds to a session in the Schelling model. The number of positive steps in a session corresponds to the number of swaps recorded, or the size of the session, S.

The distributions of S from the ABM can be compared with those from the random walk, as shown in Fig. 4. For each of the graphs shown in Fig. 4, the data from the first 800 nontrivial sessions was used. Two sample distributions from ABM are shown for the case, B=24. The lines represent the sample distributions from the random walk having a similar multiplication factor to the ABM case. If μ and T are known, the probability P can be calculated using Eq.(4) and the random walk in Eq.(5) can be defined completely. For the case μ=1, the result is shown only for the random walk, assuming T=7, because no counterpart can be found from the ABM. The distribution for μ=1 follows a power law, which can be proof that the system is in a critical condition. When μ <1.0, the scaling falls below the power law curve. Above the critical value, the distribution will have no meaning because the system will go into full segregation in the first session. The fact that the Schelling dynamics can be represented by random walks adds to our understanding of the dynamics of the Schelling model. The critical point can be represented as a balanced random walk between left-sided (sub-critical) and right-sided (super-critical) random walks. Importantly, the balanced random walk is obtained when the multiplication factor is 1.0.

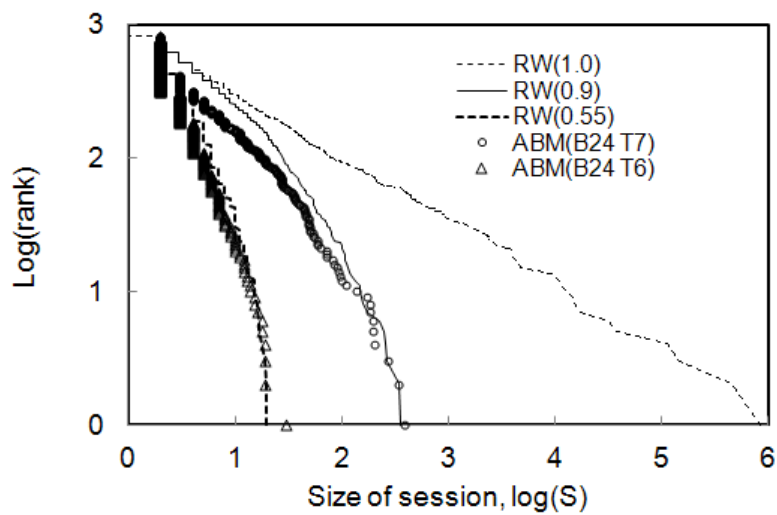

**Fig. 4.** Scaling in S. Lines represent the results from the random walk model.

## 5. Two-room model as an apatial version of the Schelling model

The condition of a chain reaction is valid only when the system satisfies the RES assumption, which can be true for earlier time frames. On the other hand, when many improving swaps accumulate in the system, the segregation will proceed and one of the RES assumptions, the randomness, will be broken. To observe the behavior of the system to full segregation, a new model reflecting such non-uniform local densities is needed. In the present study, we developed a two-room model for this purpose. Consider a system composed of two distinct rooms, called Rooms 1 and 2, as shown in Fig. 5. The two rooms represent spatially contiguous regions that are conceptually separated from each other according to the difference in local density of agents. Typically, a room will be denser in one type of agent and the other room will be denser in the other type of agent. Within a specific room, the type distribution of agents, e.g. within a neighborhood, will be assumed to be probabilistically uniform under a given density. Denote the number of each type of agent in Room1 as x and y, respectively. The agent can migrate to different rooms, and the size of the rooms can also change with increasing time steps. Initially, the system will be in a complete integration. This means that, at time step t=0 of the first session, the initial condition will be x=y=N/2. The session continues until complete segregation is obtained. Complete segregation is given as, (x,y)=(0,N/2) or (x,y)=(N/2,0). In the example cases explained below, N=10,000 was used. The two-room model, as described above, can be considered a non-spatial Schelling model because the Schelling dynamics can be reproduced without actually using the cell systems. (For details, see Appendix B)

The results from the two-room model can be compared with those from the ABM according to the number of sessions required before full segregation. For the case of the two-room model, full segregation is clearly identified, as explained above. On the other hand, for the ABM, it is difficult to judge that full segregation has been reached. A conventional method will be to use the well known indices of segregation [31]. In the present study, a measure of segregation that is similar to the dissimilarity index but modified was used (See Appendix C). The segregation index ranged from 0(well mixed) to 1.0(completely segregated). Full segregation was assumed when the index was 0.66. Figure 6 compares the number of sessions from the ABM and from the two-room model. Each of the curves shown in Fig. 6 suggests a phase transition. At subcritical values of T, the number of sessions needed to reach full segregation can go beyond the number of agents in the system. Beyond the critical demand, however, the system essentially converges to complete segregation in just one session. In this respect, the results from the ABM show that the theoretical approach described earlier is valid. The results from the two-room model coincide well for the cases of B=8 and 12. But for case B=24, the two results showed a considerable discrepancy. This suggests that some of the assumptions for the two-room model might be invalid, particularly for the case B=24. A further examination and an improvement in the two-room model will be a topic of a future study.

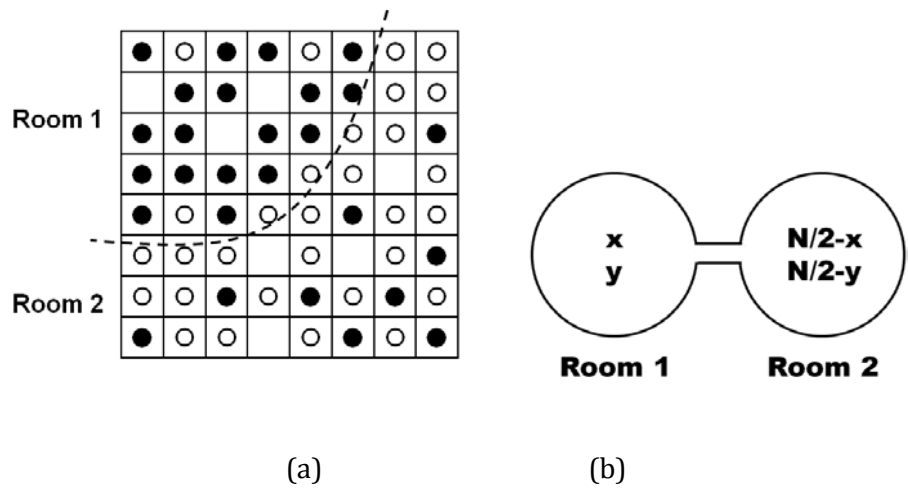

Fig.5. Two-room model. (a)Original system and (b)two-room concept.

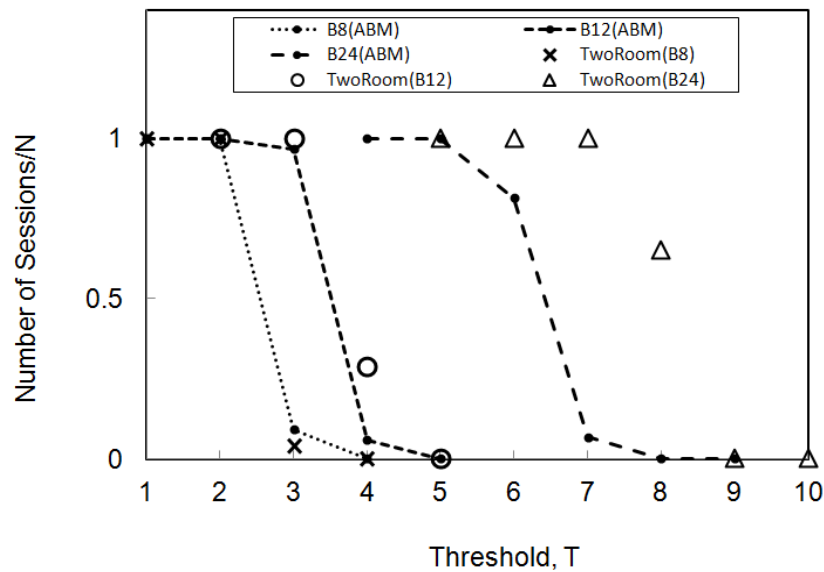

Fig. 6. Number of sessions until complete segregation. Lines were added simply for visual aid. Each of the data points is obtained by averaging 20 MC simulations.

## 6. Discussion and Conclusion

In this study, vacancies are not included in the model for simplified neighborhood conditions. If vacancies are included as in the original Schelling model, the exact results from the present study may not be valid. However, the logics used for deriving the multiplication factors may be still applicable and the two-room model could be adjusted to incorporate the vacancies. Improving and extending the present methods for a more realistic result is left as future topics of research.

A brief summary of the paper is as follows. We first proposed a new concept of multiplication factor which can be used for predicting the final outcome of the Schelling dynamics, which can be either fully segregated or not. And we devised a method for calculating the multiplication factor theoretically, making it possible to predict the final global behavior of the system without resorting to the full-scale agent-based simulation. This is claimed to be the first theoretical solution for the Schelling's critical neighborhood demand. In addition, the initial stage of the Schelling dynamics is converted into a random walk model to show the power-law behavior at the critical condition. The random walk indirectly showed that a phase transition is really involved at the critical neighborhood demand. Also we proposed an aspatial version of Schelling model, called two-room model. The numerical results from the two-room model showed a good agreement with those from the original spatial model, at least for some neighborhood topology. The aspatial version is theoretically interesting in that it is based essentially on a two-density mean-field approximation.

**References**

[1] T.C. Schelling, Dynamic Models of Segregation, J. Math. Sociology 1(1971)143-186.

[2] T.C. Schelling, Models of segregation, American Eco. Review, Papers and Proc. 59(1969) 488-493.

[3] T.C. Schelling, Micromotives and macrobehavior, W.W. Norton & Co., Inc.,1978.

[4] W.A.V. Clark, M. Fossett, Understanding the social context of the Schelling segregation model, PNAS 105 (2008) 4109-4114.

[5] L. Gauvin, J. Vannimenus, J.P. Nadal, Phase diagram of a Schelling segregation model , EPJ B 70 (2009) 293-304.

[6] A. Singh, D. Vainchtein, H. Weiss, Schelling's Segregation Model: Parameters, scaling, and aggregation, Demo. Res. 21 (2009) 341-366.

[7] S. Grauwin, E. Bertin, R. Lemoy, P. Jensen, Competition between collective and individual dynamics, PNAS 106 (2009) 20622-20626.

[8] A. Singh, D. Vainchtein, H. Weiss, Limit sets for natural extensions of Schelling's segregation model, Comm. Nonlinear Sci. Num. Simulat, 16 (2011) 2822 -2831.

[9] J. Laurie, K. Jaggi, Role of vision in neighborhood racial segregation model: A variant of the Schelling segregation model, Urban Studies 40 (2003) 2687-2704.

[10] M. Fossett, Ethnic preferences, social distance dynamics, and residential segregation: Theoretical explorations using simulation analysis, J. Math. Sociol. 30 (2006) 185-274.

[11] H. Wasserman, G. Yohe, Segregation and provision of spatially defined local public goods, Am. Economist 45 (2001) 13-24.

[12] E. Dokumaci, W.H. Sandholm,. Schelling redux: an evolutionary dynamic model of residential segregation, Unpublished manuscript, University of Wisconsin-Madison (2007).

[13] N. E. Aydinonat, Models, conjectures and exploration: an analysis of Schelling's checkerboard model of residential segregation, J. Eco. Method. 14 (2007) 429-454

[14] J. Shin, M. Fossett, Residential segretation by hill-climbing agents on the potential landscape, Adv. in Comp. Sys. 11 (2008) 875-899.

[15] I. Benenson, I. Omer, Agent based modeling of residential distribution, http://www.demogr.mpg.de/Papers/workshops/010221_paper01.pdf:1-19.

[16] A.T. Crooks, Constructing and implementing an agent-based model of residential segregation through vector GIS, Inter. J. Geo. Info. Sci. 24 (2010) 661-675.

[17] A. O'Sullivan, Schelling's model revisited: residential sorting with competitive bidding for land, Reg. Sci. and Urban Eco. 39 (2009) 397-408.

[18] A. Banos, Network effects in Schelling's model of segregation: new evidences from agent-based simulation, Environ. Plann. Des. B 39 (2012) 393-402.

[19] G. Fagiolo, M. Valente, N. J. Vriend, Segregation in networks, J. Eco. Behavior & Org. 64 (2007) 316-336.

[20] D. Vinkovic´, A. Kirman, A physical analogue of the Schelling model, PNAS 103 (2006) 19261-19265.

[21] D. Stauffer, S. Solomon , Schelling and self-organising segregation, Eur. Phys. J. B 57 (2007) 473-479.

[22] L. Dall'Asta, C. Castellano, M. Marsili, Statistical physics of the Schelling model of segregation, J. Stat. Mech. (2008) L07002.

[23] P. Sobkowicz, Simple queueing approach to segregation dynamics in Schelling model, *Preprint* 0712.3027(2008)

[24] R. Pancs, N. J. Vriend, Schelling's spatial proximity model of segregation revisited. J. Pub. Eco. 91 (2007) 1-24.

[25] S. Grauwin, F. Goffette-Nagot, P. Jensen, Dynamic models of residential segregation: An analytical solution, J. Pub. Eco. 96 (2012) 124-141.

[26] J. Zhang, Residential segregation in an all-integrationist world, J Eco. Behav. and Org. 54 (2004) 533-550.

[27] M Pollicott, H. Weiss, The Dynamics of Schelling-Type Segregation Models and a Nonlinear Graph Laplacian Variational Problem, Adv. in App. Math. 27 (2001)17-40.

[28]H. Yizhaq, B.A. Portnov, E. Meron, A mathematical model of segregation patterns in residential neighborhoods, Environ. and Plan. A 36 (2004) 149-172.

[29] J. Zhang, Tipping and residential segregation: a unified Schelling model, J. Reg. Sci. 51 (2011) 167-193.

[30]W. Feller, An introduction to probability theory and its applications, Vol. I, Random walk, Wiley Eastern (1972).

[31] D. S. Messey, N.A. Danton, The dimensions of residential segregation, Social forces 67 (1988) 281-315.

**Appendix A: Derivation of the random walk model**.

Denote the number of susceptible agents in the system at time step t by $Q_t$. If a susceptible agent is tested, there are two possibilities: he/she is unsatisfied with a probability of P or is satisfied with a probability 1-P. If satisfied, the number of the susceptible agents decreases by 1. If unsatisfied, the agent migrates and |G| susceptible agents are newly generated, resulting in a net increase in the number of susceptible agents by |G|-1. From the above discussion,

$$Q_{t+1} = \begin{cases} Q_t + |G| - 1 & \text{(with Probability P)} \\ Q_t - 1 & \text{(with Probabaility } 1 - P) \end{cases} \quad \text{(A1)}$$

Multiplying Eq.(A1) with P and using $U_t=PQ_t$ and $\mu=|G|P$ the following desired equation is obtained:

$$U_{t+1} = \begin{cases} U_t + \mu - P & \text{(with Probability P)} \\ U_t - P & \text{(with Probabaility } 1 - P) \end{cases} \quad (5)$$

**Appendix B: Two-room model**.

For the two-room model, a list **L**={$L_i$} of susceptible groups, which includes all the agents once belonging to the groups of $G_1$ and $G_2$, is maintained up to the current session. An entity in the list, $L_i$, is composed of two digits $L_i=(R_i, C_i)$ representing the room number and type of agent. At every time step, an entity in the list is selected at random and a test is performed to determine if the agent is satisfied. The agent feels satisfied if at least T same colored agent exists among the B-1 agents. Remember B-1 is used instead of B because every agent in the list already has a different colored agent that is newly swapped. The test is performed by generating the colors of B-1 agents randomly and counting the same type of agents among them. The two-room model reflects the density of the room when generating the colors of the B-1 agents. If the agent is unsatisfied, he/she will migrate to another destination and newly generated susceptible groups will be added to **L**. For the two-room model, the migration does not occur in pairs. Only the unsatisfied agent just tested migrates. Migration can occur to either of the two rooms. The probability of choosing one among the two rooms depends on the relative density of the two rooms in such a way that the agents tend to move toward a higher density room of its own color.

The values of x or y are updated if migration occurs to the other room. If the agent is satisfied, no migration will occur and only the size of L will decrease by 1. If **L** becomes empty, a session is ended and a new session can be started with a single random entity in **L**. The session continues until complete segregation is obtained. Complete segregation is given as, (x,y)=(0,N/2) or (x,y)=(N/2,0).

**Appendix C: Measure of segregation**

In the present study, the measure of segregation (M.O.S.) is defined as follows:

M.O.S.=2* (Total number of same colored neighbors)/(Total number of neighbors)-1.0

The total number means that the corresponding quantity is summed over all the agents in the system. Because the number of agents for each color is the same throughout the study, it follows that M.O.S.=0.0 when completely mixed and M.O.S.=1.0 if the two populations are completely segregated. For most of the initial RES system, the value of M.O.S. is close to 0.0 and not greater than 0.1.